\documentclass[%
 reprint,
 showkeys,
 amsmath,amssymb,
]{revtex4-2}

\usepackage{graphicx}
\usepackage{dcolumn}
\usepackage{bm}
\usepackage{physics}
\usepackage{xcolor}
\usepackage{orcidlink}

\renewcommand{\vec}[1]{\boldsymbol{\mathbf{#1}}}

\begin{document}

\title{Topology, oxidation states, and charge transport in ionic conductors}

\author{Paolo Pegolo\,\orcidlink{0000-0003-1491-8229}}%
\email{ppegolo@sissa.it}
\affiliation{SISSA---Scuola Internazionale Superiore di Studi Avanzati, 34136 Trieste, Italy}
\author{Stefano Baroni\,\orcidlink{0000-0002-3508-6663}}%
\email{baroni@sissa.it}
\affiliation{SISSA---Scuola Internazionale Superiore di Studi Avanzati, 34136 Trieste, Italy}
\affiliation{CNR---Istituto Officina dei Materiali, SISSA unit, 34136 Trieste}
\author{Federico Grasselli\,\orcidlink{0000-0003-4284-0094}}%
\email{federico.grasselli@epfl.ch}
\affiliation{ COSMO---Laboratory of Computational Science and Modelling, IMX, \'Ecole Polytechnique F\'ed\'erale de Lausanne, 1015 Lausanne, Switzerland}
\altaffiliation{Previous affiliation: SISSA---Scuola Internazionale Superiore di Studi Avanzati, 34136 Trieste, Italy}

\date{\today}
\begin{abstract}
Recent theoretical advances, based on a combination of concepts from Thouless' topological theory of adiabatic charge transport and a newly introduced gauge-invariance principle for transport coefficients, have permitted to connect (and reconcile) Faraday's picture of ionic transport---whereby each atom carries a well-defined integer charge---with a rigorous quantum description of the electronic charge-density distribution, which hardly suggests its partition into well defined atomic contributions. In this paper we review these progresses and in particular that, by relaxing some general topological conditions, charge may be transported in ionic conductors without any net ionic displacements. After reporting numerical experiments which corroborate these findings, we introduce a new connection between our topological picture and the well-known Marcus-Hush theory of electron transfer, which we are able to connect with the topology of adiabatic paths drawn by atomic trajectories. As a significant byproduct, the results reviewed here permit to classify different regimes of ionic transport according to the topological properties of the electronic structure of the conducting material. We finally report on a few recent applications to energy materials and planetary sciences.
\end{abstract}


\maketitle


\section{Introduction and outlook}\label{sec:introduction}

Ever since topology entered the field of condensed matter physics, it has proven to be a powerful tool in the classification of exotic states of matter. Many progresses have been made in the past decades in understanding how topological invariants can be computed from the electronic structure of crystalline materials, and it was recently made clear how topological effects are not an exception, but rather something that can be found in most known materials~\cite{vergniory2022all}. 
Static topological effects have been thoroughly investigated~\cite{thouless1982quantized, berry1984quantal, haldane1988model, hatsugai1993chern, kane2005z, hasan2010colloquium}; at the same time, dynamical effects were studied ever since the seminal works of Thouless and Niu~\cite{thouless1983quantization, niu1984quantised}, with the introduction of the concept of Thouless' pumps. Despite extensive works in the theory of topological quantum numbers related to the adiabatic evolution of a quantum state (see, e.g., Refs.~\onlinecite{privitera2016quantum, wauters2019localization}), the link between topology and the charge transport properties of ionic conductors has been only recently established. The starting point is a proper definition of the microscopic current~\cite{resta1992theory, resta1994macroscopic}, for which the first crucial distinction to be made is between metals and electronically insulating systems.

The electric current induced in a metal by an applied bias is carried by delocalized conduction electrons, and charge can thus flow even at fixed ions. But charge flow is not limited to metals: in ionic conductors the electrons are bound to adiabatically follow the ionic motion and no charge can be transported as long as the positions of the ions are clamped.
Nonetheless, when ions are allowed to move, charge can be macroscopically displaced. Daily-life examples range from simple salt water, to solid and liquid electrolytes employed in Li-ion batteries, or to the molten salts used as heat exchangers in power plants. Due to their large electronic bandgap, ionic conductors are in general transparent to visible light and possess a negligible fraction of ``free'' conduction electrons. 

The propensity of a material to transport charge is encoded in its electrical conductivity, $\sigma$. This  transport coefficient accounts for the relaxation of small off-equilibrium fluctuations of the electric current, as well as for the response of the latter to a small applied electric field. The modern theory of transport processes in extended systems was started by Onsager in the thirties~\cite{onsager1931reciprocal, onsager1931reciprocal2} and has been given a solid mathematical foundation back in the fifties with the Green–Kubo (GK) theory of linear response~\cite{green1952markoff, green1954markoff, kubo1957statistical, kubo1957statistical2}. This theory provides a rigorous and elegant way to cast the computation of transport coefficients into the evaluation of equilibrium microscopic fluctuations of suitably defined fluxes, thus making it accessible to equilibrium molecular dynamics (MD) simulations. In the case of the electric flux, $\vec{J}$, the associated transport coefficient, i.e., the electrical conductivity, fulfills the relation:
\begin{align}\label{eq:GK sigma}
    \sigma = \frac{V}{3 k_{\mathrm{B}} T} \int_0^{\infty} \expval{\vec{J}(t) \cdot \vec{J}(0)} \dd{t},
\end{align}
where $k_{\mathrm{B}}$ is the Boltzmann constant, $T$ is temperature, and $V$ is the volume of the system, here assumed to be isotropic, and $\langle\cdot\rangle$ denotes an equilibrium average over the initial conditions of a molecular trajectory. This formula expresses the electrical conductivity as the time-integral of the equilibrium fluctuations of the electric flux,
which amounts to saying that the transport coefficient is proportional to the product of the variance of the fluctuations of the flux times their autocorrelation time $\tau_J$~\cite{kubo2012statistical}, i.e., the time it takes for the fluctuations to regress to equilibrium:~${\sigma \propto \langle\abs*{\vec{J}}^2\rangle \tau_J}$.

In a purely classical setting~\cite{hansen2013theory}, the electric flux is just the sum of the atomic velocities times their charges:
\begin{align}
    \vec{J} = \frac{1}{V} \sum_{\ell=1}^{N} Q_{\ell} \dot{\vec{R}}_{\ell}(t), \label{eq:J-classical}
\end{align}
where $\mathbf R_\ell$ and ${Q_\ell}$ indicate the position and the charge of the $\ell$th atom, respectively, $V$ the system's volume, and $N$ the number of atoms.
When a quantum mechanical picture of the interatomic forces is adopted, the situation is not nearly as clear: in this case, the ionic cores do possess a well-defined positive integer charge, whereas the continuous nature of the electronic charge-density distribution poses a challenge to a mathematically rigorous and physically meaningful identification of the valence contribution to the ionic charges. The impasse is resolved by realizing that the electric flux is the time-derivative of the macroscopic polarization, ${\vec{J}=\dot{\vec{P}}}$, and by evaluating this time derivative using the chain rule. In order to do so, one defines the atomic Born effective charges~\cite{ghosez1998dynamical,resta1994macroscopic}, 
\begin{equation} \label{eq:BornZ*}
    \bm{\mathsf{Z}}^*_{\ell} \doteq \frac{\partial\vec{P}}{\partial \vec{R}_\ell},
\end{equation}
where $\vec{R}_\ell$ is the position of the $\ell$th atom, and one concludes that the electric flux can be cast in the same form as in the classical case, provided the time-independent, integer-valued, scalar ionic charges are replaced by the time-dependent, real-valued tensor, Born effective charges~\cite{grasselli2019topological, french2011dynamical}:
\begin{align} \label{eq:J Born}
    \vec{J}(t) = \frac{1}{V} \sum_{\ell=1}^N \bm{\mathsf{Z}}^*_{\ell}(t) \cdot \dot{\vec{R}}_{\ell}(t).
\end{align}
It is immediate to realize that, since the effective charges are in general nonzero, due to the local variations of the atomic or molecular dipoles, so are the fluctuations of the flux, even when the system is made of neutral molecules, such as in the case of, e.g., pure, undissociated water. The question then naturally arises as to how come nonvanishing flux fluctuations give rise to the vanishing electrical conductivity of pure water or other mixtures of neutral moieties.

In the literature, one can find ab initio simulations employing both the rigorous quantum mechanical definition of the electric flux of Equation~\eqref{eq:J Born}~\cite{french2011dynamical, sun2012direct, zhao2016first}, as well as others where integer charges are assigned to atoms in motion~\cite{french2010diffusion, nilsson2016ionic, marcolongo2017ionic}. Both approaches lead to satisfying results: while in the former case this is, of course, to be expected, since it makes use of the exact formul\ae, the latter could be thought reasonable in strongly ionic cases---where electronic polarization effects are allegedly small and effective charge tensors could possibly be \emph{approximated} to integer multiples of the identity---but questions might arise in all the other cases, where covalent effects are not negligible~\cite{resta2021faraday}. Instead, for reasons that have been baffling until they were recently demystified, integer charges were found to work even in nonionic systems. A notable example is provided by Ref.~\onlinecite{french2011dynamical}, where the electrical transport properties of partially dissociated H\textsubscript{2}O---a phase of water occurring at high-temperature and high-pressure conditions---are investigated. In that paper, the authors noticed that \emph{interestingly, the use of predefined constant charges can yield the same conductivity as is found with the fully time-dependent charge tensors}. Even more interestingly, the values of those charges are what chemical intuitions would suggest for the oxidation states (OSs) of hydrogen (${+1}$) and oxygen (${-2}$). As a coronation of this, the use of other possibly significant scalar quantities, such as Bader~\cite{bader1990atoms} or Mulliken~\cite{mulliken1955electronic} charges, the average of the diagonal of the Born effective charges over the constituents of the system~\cite{sun2015the}, or even the effective charges coming from the electrostatics in empirical force fields~\cite{dufils2018properties, dufils2020a}---despite having been effectual on occasions~\cite{goldman2009ab, sun2015the, rozsa2018ab}---is in principle wrong, and does not work in general~\cite{french2011dynamical}. 

In this Review, we analyze the existing literature on the subject~\cite{grasselli2019topological, pegolo2020oxidation, resta2021faraday} to highlight that these apparent coincidences are by no means such, but are instead the deep manifestation of the topological properties of the insulating state, which is maintained as the ions are allowed to diffuse adiabatically through macroscopic distances. By leveraging Thouless' quantization of particle transport~\cite{thouless1983quantization}, we first show that, under suitable topological conditions, every atom in an ionic conductor can be characterized by an integer \emph{topological charge}, which is the same for all the atoms of a same chemical species, and whose properties closely echo those of the atomic OSs~\cite{grasselli2019topological}. The combination of this finding with a recently introduced \emph{gauge-invariance} principle for transport coefficients allows us to shed light onto the alleged coincidences reported above. It was shown that, whenever the topology of the insulating state permits to define species-dependent OSs, the electrical conductivity resulting from GK formula, Equation~\eqref{eq:GK sigma}, when the electric flux is defined in terms of these charges is the same that would result from the use of the Born effective charges~\cite{grasselli2019topological}.

The theory was also extended to two other different situations where a unique, species-dependent attribution of OSs is not possible, giving rise to charge transfer mechanisms that are not accompanied by a net ionic displacement. 
The first such situation occurs when not all the atoms of the same chemical species have the same OS~\cite{jiang2012rigorous}.
In the second, OSs are not well-defined at all, which results in a nontrivial ionic transport regime, mediated by the adiabatic motion of localized electronic charge not bound to any specific atom~\cite{pegolo2020oxidation}. The topology of the system under consideration determines which of these cases occurs. For these charge-transfer mechanisms, we  establish a new connection with the Marcus-Hush theory of electron transfer~\cite{marcus1956on, hush1958adiabatic}, providing a common picture for charge-transfer reactions and the conditions which allow charge to flow without ionic mass transport.

The first Sections of this work are devoted to reviewing recent theoretical advances in the role of topology in the theory of charge transport in ionic conductors. After a brief summary of the theory of ionic transport in electronically insulating liquids, outlined in Section~\ref{sec:ionic transport}, the deep connection between charge transport and the theory of polarization will be discussed in Section~\ref{sec:charge transport theory of polarization}~\cite{grasselli2019topological, resta2021faraday}. With these tools, and under suitable conditions on the topology of the system, a rigorous definition of OSs will be given, together with its effect on the transport properties of ionic conductors, in Section~\ref{sec:topological oss}~\cite{grasselli2019topological}. The following Section \ref{sec:wannier} is devoted to the case of independent electrons, where a picture in terms of Wannier Functions (WFs) is particularly insightful~\cite{marzari1997maximally}. Then, in Section~\ref{sec:breakdown of SA}, the topological conditions that allowed to define OSs will be relaxed, and the consequences of this will be extensively analyzed~\cite{jiang2012rigorous, pegolo2020oxidation}. In Section~\ref{sec:charge transfer}, we present a newly introduced link between the topological description of charge transport and the Marcus-Hush theory of charge transfer reactions. In Section~\ref{sec:applications}, we report on some practical applications, while Section~\ref{sec:conclusions} contains our conclusions.

\section{Charge transport in ionic conductors}
\label{sec:ionic transport}

A material can be thought of as a collection of classical nuclei and quantum electrons. Ionic transport in an electrolytic cell is the transfer of a charged nucleus from one electrode to the other after it has traversed the electrolyte solution, driven by an electric field. Consequently, electrons are pumped in the same direction through the battery put between the two electrodes to maintain charge neutrality in the cell.

To model charge transport in a closed quantum system, one must adopt periodic boundary conditions (PBCs), as they are the only ones able to sustain a steady-state flux~\cite{resta2017the}. The constituents of the system are therefore supposed to be contained in a cube of side $L$ and volume $V=L^3$, and whose opposing faces are identified because of PBCs. The three coordinates $x$, $y$, and $z$, can thus be mapped to a triplet of angles $\bm{\varphi} = 2 \pi (x, y, z) / L$.

The central quantity in ionic transport is the electrical (ionic) conductivity, $\sigma$, expressed, in the GK formalism, by Equation~\eqref{eq:GK sigma}. 
In a purely classical atomistic model of a liquid, the expression of the electric flux is given by Equation~\eqref{eq:J-classical}.
The values of $\{Q_{\ell}\}$ are usually chosen to reproduce in some way the physical properties of the system of interest, depending on the problem at hand. The values of the (static) atomic charges range from integer numbers in the case of ionic compounds to real numbers in the case, e.g., of partial charges in molecules~\cite{gross2002comparison, heinz2004atomic}. While any purposely chosen value might make sense from the point of view of the electrostatics of the particular model, charge transport requires the charges to be \emph{integers}, as it will be shown in the following; using real-valued atomic charges evaluated from static calculations would lead to an error of principle when computing the conductivity of an electronically insulating material.

When the quantum nature of electrons is taken into account one needs to resort to Equation~\eqref{eq:J Born}, where Born effective charge tensors replace the scalar atomic charges. The evaluation of the GK formula, Equation~\eqref{eq:GK sigma}, from first principles requires the computation of Born effective charges along a MD trajectory, using either linear-response theory~\cite{baroni2001phonons} or a Berry-phase approach~\cite{vanderbilt2018berry}.

In the case of ionic conductors in an electrolytic cell, however, it is known since Faraday's times that when $N$ members of a given chemical species (be they atomic or molecular moieties) pass from one electrode to the other, the charge that is transported is a specific integer multiple of the elementary charge, $e$, times $N$~\cite{faraday1834vi, resta2021faraday}. In this way, it is experimentally possible to measure an integer charge pertaining to each member of the chemical species at hand. It is important to notice that the measurement is intrinsically dynamical, in that it requires the motion of nuclei across a macroscopic distance, given by the size of the electrolytic cell. The integer charge measured by Faraday is an example of {OS}. In chemistry, OSs are integer numbers widely used to describe redox reactions, electrolysis, and many electrochemical processes. In spite of their fundamental nature, they have long eluded a rigorous quantum-mechanical interpretation. In fact, they are usually determined according to an agreed set of rules~\cite{walsh2018oxidation, raebiger2009oxidation}, their simplest---and official---definition being, for an atom, \emph{the charge of this atom after ionic approximation of its heteronuclear bonds} (verbatim from the IUPAC Gold Book~\cite{mcnaught1997compendium}). Yet, being measurable quantities, at least in the electrolytic cell setup, they are expected to have a rigorous quantum-mechanical definition which must reflect the fact that ionic conductors carry integer charges over macroscopic distances.

\section{Charge transport and the theory of polarization}\label{sec:charge transport theory of polarization}

\subsection{Single-point Berry phase}

In a completely general way, the modern theory of polarization~\cite{vanderbilt1993electric, vanderbilt2018berry} allows one to write the electric current as an elegant expression that involves the time derivative of a phase angle. We shall adopt the notation of Ref.~\onlinecite{resta2021faraday}, where the many-body generalization of the theory introduced in Ref.~\onlinecite{grasselli2019topological} is presented. This allows to move from an independent-electron picture to a fully many-body one, in a way analogous to how Ref.~\onlinecite{niu1984quantised} upgraded the original Thouless' paper~\cite{thouless1983quantization} to its many-body formulation. In this way, the results we present in this Subsection are exact in the adiabatic limit at any level of theory, and their applicability is only limited by the approximations adopted when explicit numerical calculations are done. Let $\ket{\Psi(t)}$ be the instantaneous adiabatic ground state of a quantum system with $N$ nuclei and $N_{\mathrm{el}}$ electrons. The adiabatic electric flux that takes into account PBCs is~\cite{resta1998quantum,resta2000manifestations,resta2021faraday}
\begin{align}
    \label{eq:J im log}
    \vec{J}(t) &= \frac{e}{2 \pi L^2} \dv{}{t} \lim_{L \to \infty} \Im \log \bm{\mathfrak{z}}(t), \\
    \label{eq:zeta definition}
    \mathfrak{z}_{\alpha}(t) &= \mel{\Psi(t)}{\mathrm{e}^{i \frac{2\pi}{L}\left(\sum_{\ell=1}^{N} Z_{\ell} \hat{R}_{\ell,\alpha}-\sum_{j=1}^{N_{\mathrm{el}}}\hat{r}_{j,\alpha}\right)}}{\Psi(t)},
\end{align}
where ${\sum_{j=1}^{N_{\mathrm{el}}}\hat{\vec{r}}_{j}}$ is the position operator of the electrons, and each nuclear species, $S$, characterized by an atomic number $Z_S$, has a position operator given by ${\sum_{\ell=1}^{N_{\mathrm{S}}}\hat{\vec{R}}_{\ell}}$, with ${\sum_{S} N_S = N}$. The index ${\alpha = x,y,z}$ indicates the Cartesian component. Since the exponential operator in Equation~\eqref{eq:zeta definition} is unitary, its expectation value is a complex number of at most unit modulus. The sheet of the $\Im\log$ function, which features a branch-cut singularity in the complex plane, is chosen so that $\vec{J}$ is continuous in time.
A triplet of phase angles,
\begin{align}
    \bm{\gamma}(t) = \lim_{L\to\infty}\Im \log \bm{\mathfrak{z}}(t),
\end{align}
is thus defined, and its electronic component is known in the literature as the \emph{single-point Berry phase}~\cite{resta1998quantum}. In terms of $\bm{\gamma}(t)$, the electric flux takes the compact form~\cite{resta2021faraday}
\begin{align}\label{eq:J as derivative of gamma}
    \vec{J}(t) = \frac{e}{2 \pi L^2} \dot{\bm{\gamma}}(t).
\end{align}
The dipole displaced along a trajectory of the system, i.e.,~${\bm{\Delta\mu}(t)=L^3\int_0^t \vec{J}(t^{\prime})\dd{t^{\prime}}}$, is thus given by
\begin{align}\label{eq:Delta mu as a phase difference}
    \bm{\Delta\mu}(t) = e L \frac{\bm{\Delta \gamma}(t)}{2\pi},
\end{align}
with ${\bm{\Delta\gamma}(t)=\bm{\gamma}(t)-\bm{\gamma}(0)}$. 
Equation~\eqref{eq:Delta mu as a phase difference} says that, as a consequence of the adiabatic evolution of the system, the dipole displaced is proportional to a vector of accumulated phase angles---one component per Cartesian direction.

Pictorially, $\Delta\gamma_\alpha(t)$ represents the angle spanned by $\mathfrak{z}_{\alpha}$ in the complex plane, and $\Delta\gamma_\alpha(t)/(2\pi)$ is the number (in general non-integer) of rotations that $\mathfrak{z}_{\alpha}$ makes around the origin. 
If the adiabatic ground state returns to itself after a time $\tau$, ${\ket{\Psi(\tau)}=\ket{\Psi(0)}}$, the system's evolution is cyclic, and the accumulated angle is an integer multiple of $2\pi$; therefore, the transported charge is an integer multiple of the elementary charge. 
\begin{figure}[htb]
    \centering
    \includegraphics[width=\columnwidth]{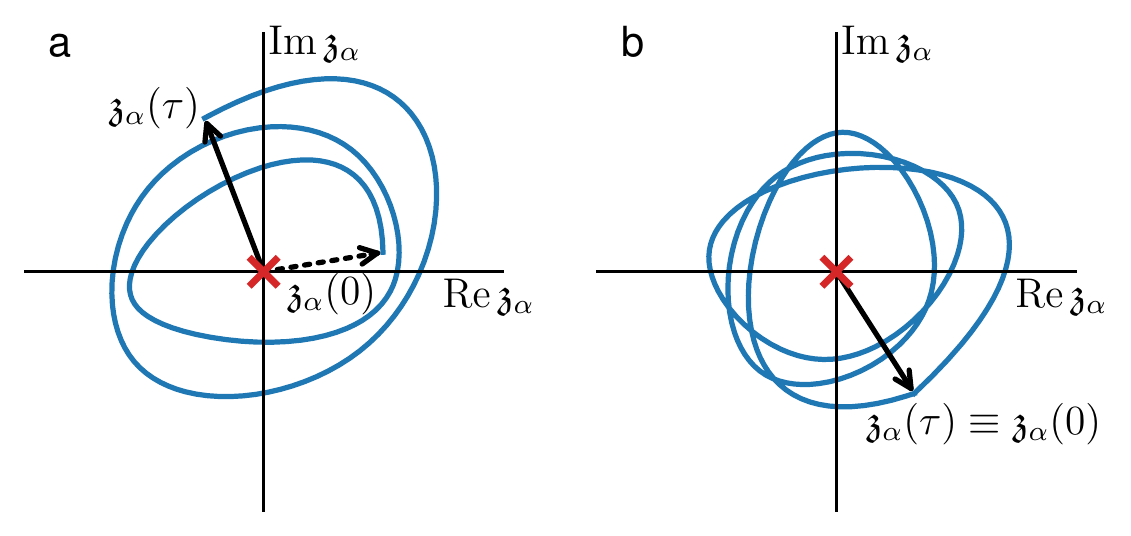}
    \caption{Sketch of the path drawn by $\mathfrak{z}_{\alpha}$ in the complex plane during the time-evolution of the system. (a) For a general nuclear trajectory, the path is open. (b) If the initial and final configurations of the nuclei are the same, then also $\bm{\mathfrak{z}}$ returns to itself: the path in the complex plane is closed.}
    \label{fig:zeta winding}
\end{figure}
The two scenarios---the case of a generic trajectory, and the one where the trajectory is cyclic---are visualized in Figure~\ref{fig:zeta winding}: on the one hand (a), a general dynamics of the system traces open paths in the complex plane; on the other (b), the same initial and final adiabatic ground states imply that also $\bm{\mathfrak{z}}$ returns to itself. The latter situation defines the number of windings around the origin of the complex plane as a topological invariant, since deforming the path traced by $\bm{\mathfrak{z}}$ does not change its \emph{winding number}, provided that the path itself does not \emph{cross} the origin. This condition, i.e.,~${\abs{\mathfrak{z}_\alpha(t)} \neq 0}$, for all $\alpha$ and for each $t$, amounts to saying that the system stays insulating during the entire dynamics~\cite{resta1999electron, resta2021faraday}. In fact, according to the celebrated theory of Resta and Sorella, $\mathfrak{z}$ is related to whether the system is insulating or metallic, its modulus tending to $1$ from below in insulators, and to $0$ from above in metals, up to the leading order in $1/L$~\cite{resta2018theory}.

\begin{figure*}[htb]
    \centering
    \includegraphics[width=2\columnwidth]{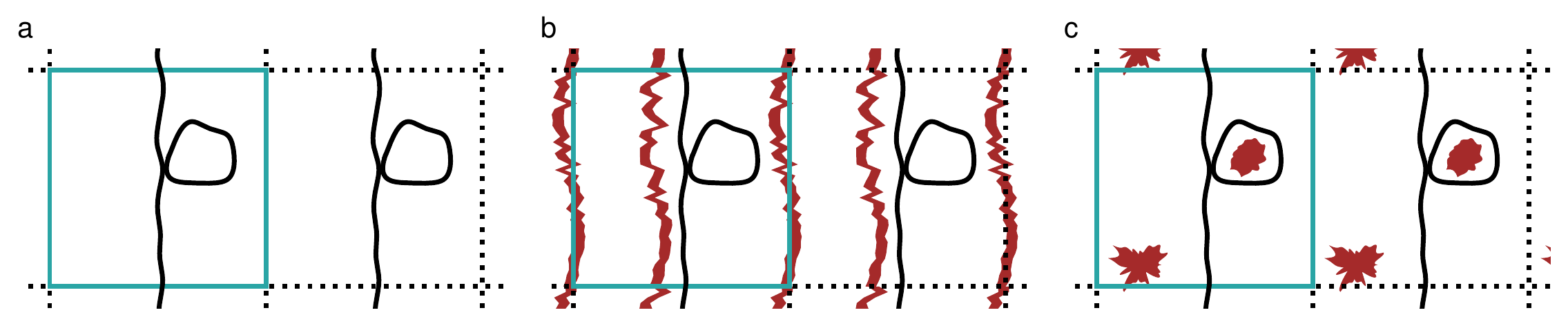}
    \caption{Different types of ACSs, according to the metallic regions they feature. The central cell is framed in light blue, while the dotted lines represent the sides of one of its periodic replicas. Metallic regions are depicted in dark red. The black lines are paths in the ACS. (a) An ACS where SA holds everywhere. (b) Metallic walls surround an adiabatic domain, where SA holds locally. (c) Here, SA is broken, in that there exist trivial loops that encircle metallic regions.}
    \label{fig:ACS zoology}
\end{figure*}

\subsection{Born-Oppenheimer approximation}
Let us now focus on a model liquid where \emph{the nuclei are classical particles, while electrons maintain their quantum nature}. The adiabatic limit here takes the form of the Born-Oppenheimer approximation: at each time, the electronic wavefunction obeys a time-independent Schr\"odinger equation whose one-body potential term is determined by the instantaneous nuclear coordinates, $\{\vec{R}_{\ell}(t)\}$. Equation~\eqref{eq:J im log} and~\eqref{eq:zeta definition} keep their form,  but $\ket{\Psi(t)}$ is replaced by the purely electronic ground state, and the operators $\hat{\vec{R}}_{\ell}$ are replaced by the classical nuclear positions $\vec{R}_{\ell}$.
Under these conditions, the electric flux is given by Equation~\eqref{eq:J Born}, where the fact that the polarization depends on time only through the nuclear coordinates is made explicit. In the ${L \to \infty}$ limit, Equation~\eqref{eq:J im log} is equivalent to Equation~\eqref{eq:J Born}, but only the latter is manifestly additive in the nuclear contributions. It is then expedient to define $3N$ quantities, $\gamma_{\ell,\alpha}$, one for each nucleus and Cartesian direction, such that~\cite{resta2021faraday}:
\begin{align}
    \label{eq:dot gamma per nucleus} \dot{\gamma}_{\ell, \alpha}(t) &= \frac{2 \pi}{L} \sum_{\beta} \mathsf{Z}^{*}_{\ell, \alpha \beta}(t) \dot{R}_{\ell, \beta}(t), \\
    \label{eq:gamma per nucleus} \gamma_{\ell, \alpha}(t) &= \int_0^t \dot{\gamma}_{\ell, \alpha}(t^{\prime}) \dd{t^{\prime}}.
\end{align}
In this way, the total accumulated phase angle is decomposed into a sum of contributions from each nucleus, i.e.,
\begin{align}\label{eq:gamma as sum over nuclei}
    \bm{\Delta\gamma}(t) = \sum_{\ell=1}^N \bm{\gamma}_{\ell}(t).
\end{align}

The emergence of the topological invariants described above has deep consequences on the charge transport properties of ionic conductors and, in particular, on the electrical conductivity.
The electrical conductivity can be written in a form which is equivalent to the GK formula, Equation~\eqref{eq:GK sigma}, i.e.,~the Einstein-Helfand (EH) relation~\cite{helfand1960transport}:
\begin{align}\label{eq:einstein-helfand sigma}
    \sigma = \frac{1}{3 L^3 k_\mathrm{B} T} \lim_{t \to \infty} \frac{\expval{\abs{\bm{\Delta \mu}(t)}^2}}{2 t},
\end{align}
where $\bm{\Delta\mu}$ is the displaced dipole defined in Eq.~\eqref{eq:Delta mu as a phase difference}. Any two expressions of $\bm{\Delta\mu}$ that differ by a bounded vector result in the same electrical conductivity: the most general form of this statement, which holds for any transport coefficient, takes the name of \emph{gauge invariance of transport coefficients}~\cite{marcolongo2015microscopic, bertossa2019theory, grasselli2019topological, baroni2020heat, Grasselli2021invariance}. Simply put, gauge invariance says that transport coefficients are largely independent of the detailed expression of the local representation of the conserved quantity, be it mass, energy or, as in the present case, charge. By leveraging this concept, it can be shown that, under suitable hypotheses on the topology of the system, it is possible to assign integer, constant OSs, $\{q_{\ell}\}$, to each one of the $N$ atoms, and that the resulting electrical conductivity, $\sigma^{\prime}$, given by Equation~\eqref{eq:einstein-helfand sigma} with the dipole displacement defined by
\begin{align}\label{eq:Delta mu topological charges}
    \bm{\Delta \mu}^{\prime}(t) = \sum_{\ell=1}^{N} e\, q_{\ell} \int_0^t \dot{\vec{R}}_{\ell}(t^{\prime}) \dd{t^{\prime}},
\end{align}
has the same value as the fully quantum-mechanical definition based on Equation~\eqref{eq:J Born}, which employs time-dependent Born effective charge tensors. In formul\ae:
\begin{align}\label{eq:GB theorem}
    \lim_{t \to \infty} \frac{\expval{\abs{\bm{\Delta \mu}(t)}^2}}{2 t} = \lim_{t \to \infty} \frac{\expval{\abs{\bm{\Delta \mu}^{\prime}(t)}^2}}{2 t},
\end{align}
which is to say $\sigma=\sigma^{\prime}$~\cite{grasselli2019topological}.
This equation is valid in the thermodynamic limit ${L,N \to \infty}$ at fixed density.

\subsection{Paths and topology}\label{sec:topology of paths}
Before proving Equation~\eqref{eq:GB theorem} in the next section, we make some considerations on the topology of the paths generated by the adiabatic dynamics of the nuclei, as these aspects have crucial repercussions on the charge transport properties of the system. The space of all the coordinates of the nuclei---namely, the atomic configuration space (ACS)---is topologically equivalent to a $3N$-dimensional torus, since we adopt PBCs in all the three Cartesian coordinates of each of the $N$ nuclei, independently. We are interested in those paths whose endpoints are a periodic image of one another, since it is for these that the adiabatic ground state returns to itself and the theorems on quantization of charge can be invoked. We focus on \emph{adiabatic paths}, i.e., paths in the ACS that never cross a metallic region, so to keep the evolution adiabatic. In this light, it is expedient to define the \emph{adiabatic space} as the ACS deprived of the regions where the system is metallic. Different types of charge transport in insulators arise, depending on whether the classification of paths of the adiabatic space coincides with that of the total ACS, and, if they differ, how they do so. We outline the characteristics of these paths with reference to the situations sketched in Figure~\ref{fig:ACS zoology}.
If each path whose endpoints are replicas of one another can be uniquely specified, up to deformations which do not imply cuts or ``exiting the space'', by the tuple of $3N$ integers, ${\vec{n}=(n_{1x},n_{1y},\ldots,n_{Nz})}$, representing the number of cells spanned by each atom in each Cartesian direction, then \emph{strong adiabaticity} (SA) holds in the adiabatic space~\cite{grasselli2019topological, pegolo2020oxidation}. This means that a trivial loop in the ACS (i.e., a path whose endpoints belong to the same cell and coincide, and is thus characterized by $\vec{n} = \bm{0}$) is also a trivial loop in the adiabatic space, and can be shrunk to a point without crossing any metallic region. The easiest situation is whenever there are no metallic regions for any configuration of the nuclei: this is the case sketched in a two-dimensional ACS in Figure~\ref{fig:ACS zoology}a. Notice that, in higher dimensions, the complete insulation of the ACS is not necessary for SA: for instance, removing a metallic point (or even a ball) from a 3-dimensional torus preserves SA of the resulting adiabatic space. 

In other cases, there might be metallic regions in the ACS that partition the full adiabatic space into disconnected adiabatic \emph{domains}~\cite{pegolo2020oxidation}, playing the role of ``walls'' that the system would need to cross if it were to pass from one adiabatic domain to the other. The classifications of paths on the ACS and on the adiabatic space no longer coincide, since paths with some $\vec{n}$ are not allowed in the entire adiabatic space, as they would imply crossing a metallic region. Nevertheless, if all the adiabatic trivial loops can still be shrunk to a point without closing the gap, the same considerations made above for the case where SA holds everywhere are still valid \emph{within a given adiabatic domain}~\cite{pegolo2020oxidation}. This situation is sketched in Figure~\ref{fig:ACS zoology}b. Here, SA holds in each adiabatic domain, but not on the whole ACS, due to the presence of metallic walls.

The situation is instead totally different whenever there exist adiabatic trivial loops in the ACS which, despite entirely belonging to the adiabatic space, cannot be shrunk to a point without crossing a metallic region~\cite{grasselli2019topological,pegolo2020oxidation}. Here, adiabatic paths that are connected are not necessarily equivalent to one another, which implies that two paths characterized by the same tuple $\vec{n}$ may be topologically different. This is shown in Figure~\ref{fig:ACS zoology}c. Here, SA does not hold anymore. 

Whether SA holds or not has profound consequences on the definition of OSs and on the transport properties of insulating systems. In the following Section we shall see that only when SA is not dropped the quantum theory of charge transport can be reconciled with Faraday's description, where each atom can be endowed with a well-defined integer charge, and Equation~\eqref{eq:GB theorem} holds~\cite{grasselli2019topological, pegolo2020oxidation}.

\begin{figure}[htb]
    \centering
    \includegraphics[width=0.85\columnwidth]{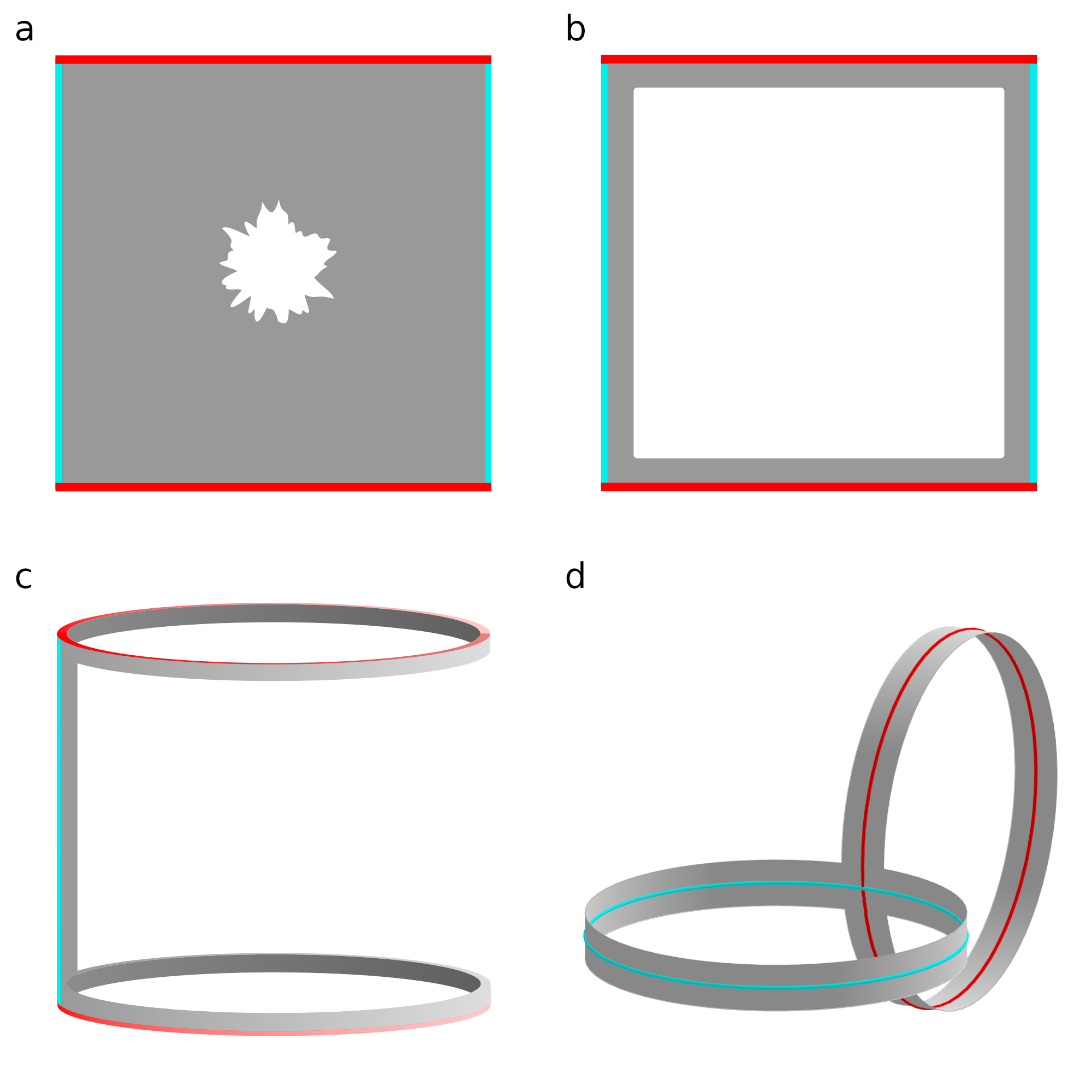}
    \caption{A puntured 2-torus retracts to a figure eight. (a) The punctured 2-torus is represented as a square whose opposite sides are identified (as in PBCs) with a hole in it. (b) Since we are interested to the topology of the manifold, the hole can be continuously expanded. (c) The blue sides are glued together. (d) Finally, the red sides are glued together. The resulting manifold, whose fundamental group is no longer Abelian, can be retracted to a ``figure eight'', $\textcolor{red}{\bigcirc}\!\!\cdot\!\!\textcolor{cyan}{\bigcirc}$. Its fundamental group, $\mathbb{Z}*\mathbb{Z}$ differs from that of the 2-torus, $\mathbb{Z}\times\mathbb{Z}$. Nevertheless, its abelianization, i.e., the first homology class, is $\mathbb{Z}\times\mathbb{Z}$, and the ``elementary loops'' $\textcolor{red}{\bigcirc}$ and $\textcolor{cyan}{\bigcirc}$ coincide with those of the torus. In this way, charge transport properties are not altered by a single puncture of the 2-torus.}
    \label{fig:punctured torus}
\end{figure}

\begin{figure}[htb]
    \centering
    \includegraphics[width=0.7\columnwidth]{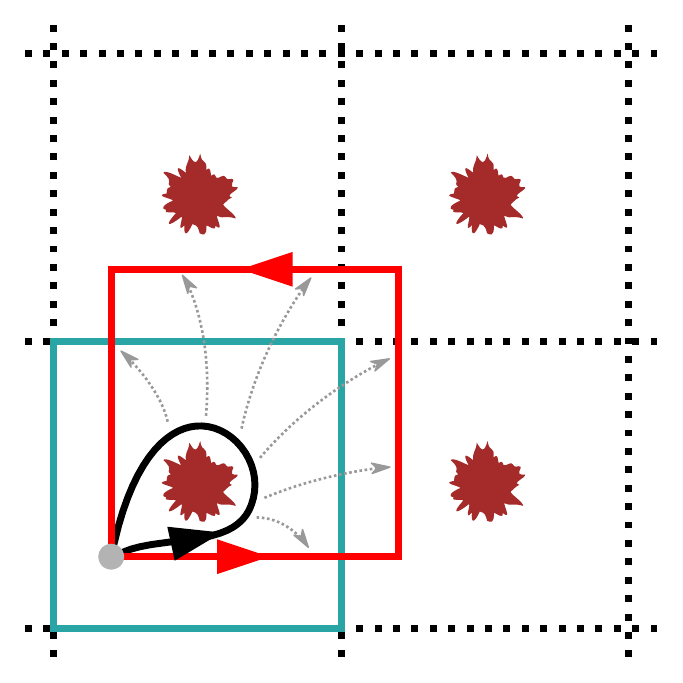}
    \caption{Sketch of a 2-dimensional ACS isomorphic to a punctured torus. A single hole is sufficient to change the topological properties of the manifold, as it changes, e.g., its fundamental group (see Figure~\ref{fig:punctured torus}). Nonetheless, the black path can be continuously deformed (the deformation map is represented by the gray, dotted, arrows) to the red one without closing the gap: it is evident that the charge transported along the latter is zero, since the charges transported along each side of the square cancel out with those transported along the opposite side, which is just the same path traversed in reverse.}
    \label{fig:homology}
\end{figure}

\subsection{Homotopy and homology in charge transport}\label{sec:homology}
Two given paths are considered topologically equivalent, or \emph{homotopic}, if they can be deformed one into another without cuts nor exiting the manifold they belong to~\cite{hatcher2001algebraic}. 
Closed paths starting and ending at the same (base) point---i.e., loops---can be classified accordingly, and the set of the equivalence classes of loops under homotopy, together with the operation of concatenation, is called the fundamental group of the manifold, which incorporates information on the presence of important features (like ``holes'' or ``handles'') of the manifold~\cite{hatcher2001algebraic}. The fundamental group can be commutative (\emph{Abelian}) or not. For instance, the fundamental group of the ($3N$-dimensional) torus is Abelian, since it is isomorphic to $\mathbb{Z}^{3N}$.
Nonetheless, when we ``puncture'' it (as it happens when we remove metallic regions from the ACS), the fundamental group of the new manifold (that we named \emph{adiabatic space} in Section~\ref{sec:topology of paths}) may change~\cite{hatcher2001algebraic}, which is exactly what happens when SA is broken~\cite{pegolo2020oxidation}. This new fundamental group may be no longer Abelian. In spite of this, the total transported charge is expressed as a line integral that can be split into the \emph{commutative} sum of integrals, each computed over a piece of the original path.
Therefore, it is not strictly necessary that two loops are homotopic to transport the same integer charge. In fact, it is sufficient that they are \emph{homologous}, i.e., that they share the same, possibly repeated, ``elementary loops'', with no further restriction on the sequence in which these are taken. 
More rigorously, to understand if two paths transport the same charge, one must inspect whether they are represented by the same element of the so-called \emph{first homology group}, which is the ``abelianization'' of the fundamental group~\cite{hatcher2001algebraic}. The use of concepts from homology theory is also used in topological data science applied to materials science~\cite{nakamura2015persistent,hiraoka2016hierarchical,ichinomiya2017persistent,buchet2018persistent}. An example of the difference between homotopy and homology is shown in Figures~\ref{fig:punctured torus} and~\ref{fig:homology}; a single metallic region in a 2-dimensional ACS in PBCs is enough to change its topology from a simple torus to a ``figure-eight'' (see Figure~\ref{fig:punctured torus}): the fundamental group changes from the Cartesian product ${\mathbb{Z} \times \mathbb{Z}}$, which is Abelian, to the free product ${\mathbb{Z} * \mathbb{Z}}$, which is not Abelian~\footnote{$\mathbb{Z}\times\mathbb{Z}$ or, equivalently, $\mathbb{Z}^2$, is the group of ordered pairs with the commutative operation of sum. It is Abelian, since ${(n,m) + (p,q) \equiv (p+n, q+m) = (p,q) + (n,m)}$. 
Instead, the free product ${\mathbb{Z} * \mathbb{Z}}$ is for instance represented by the group of two letters (and their inverse), with the non-commutative operation of justapposition, where order matters, as in natural language, where being ``OK'' differs from being ``KO''!}. 
Nonetheless, the first homology group remains the same, and the transported charge along a path that encircles the metallic region is zero, as it can be deformed to a squared path whose opposing faces transport charges which manifestly cancel out with one another, as shown in Figure~\ref{fig:homology}. The black loop and any trivial loop that shrinks to the gray point are homologous, despite they are not homotopic, as they cannot be continuously deformed into one another without crossing the metallic region. Notice that two points need to be removed from a 2-torus to change the first homology group, and make it non trivial from the transport point of view, as reported below for the linear H$_3^+$ cation discussed in Ref.~\onlinecite{pegolo2020oxidation}.

\section{Topological foundation of the oxidation states}\label{sec:topological oss}

Under the condition of SA in the whole ACS, let us now characterize the integer charges appearing in Equation~\eqref{eq:Delta mu topological charges}. We follow a \emph{Gedankenexperiment}, proposed for the first time by Pendry and Hodges~\cite{pendry1984the}. We displace  a single nucleus, say the $\ell$th, from its initial position to the same position in an adjacent replica cell. To perform this task, the other ions are allowed to move out of the way provided their final positions coincide with the initial ones and the electronic gap never closes along the path~\cite{grasselli2019topological, resta2021faraday}. Thouless' theorem~\cite{thouless1983quantization} (and the many-body generalization thereof~\cite{niu1984quantised}) ensures that the total charge displaced along this trajectory is an integer multiple of the elementary charge: as it is also expressed by Equation~\eqref{eq:Delta mu as a phase difference} in the case of a cyclic path, this integer number is the winding number of $\mathfrak{z}_{\alpha}$ around the origin of the complex plane (with $\alpha$ the Cartesian direction around which the nucleus has moved). Let $q_{\ell, \alpha}$ be this number, which we call the \emph{topological charge} of the $\ell$th atom along direction $\alpha$. We remark that, for any \emph{static} configuration taken from such artificial evolution, these topological charges cannot be defined. A necessary ingredient for their characterization is the \emph{dynamic} displacement of the selected atom. 

An operation of this kind was used by Jiang \textit{et al.}~\cite{jiang2012rigorous} in the case of crystalline solids, where the system is periodic and lattice translations bear physical meaning. More generally, the values of the topological charges are independent of the macroscopic size of the system closed under PBCs. This validates the procedure also for disordered systems~\cite{grasselli2019topological}. Moreover, in an electrolytic cell, it is possible to map the physical system to a spatially periodic one, as the electrodes are connected to one another by a wire~\cite{resta2021faraday}.

In order to establish an equivalence between topological charges and OSs, three features ascribed to the latter must occur in $\{q_{\ell,\alpha}\}$: \textit{i}) the topological charges have to be path-independent, \textit{ii}) their value must coincide for equivalent atoms, and \textit{iii}) they must be \emph{isotropic}, i.e.,~independent of the specific Cartesian direction by which the atom is moved. The first condition is easily proved under SA: consider two different paths having the same endpoints; by the very definition of SA, two such paths can be deformed into one another without ever leaving the insulating state, i.e.,~without any component of $\bm{\mathfrak{z}}$ passing through the origin of the complex plane. The second condition can be proved by invoking the additivity of integrals~\cite{grasselli2019topological, resta2021faraday}: take any two nuclei, labeled $\ell^\prime$ and $\ell^{\prime\prime}$, of the same species $S$, i.e.,~with the same atomic number, $Z$; the phases, separately accumulated by each of them in a cyclic path that brings them to their replica in the adjacent position along Cartesian direction $\alpha$, yield topological charges $q_{\ell^\prime,\alpha}$ and $q_{\ell^{\prime\prime},\alpha}$. Since SA holds, the initial positions of the two nuclei can be exchanged without crossing a metallic state. Let the nucleus $\ell^\prime$ \textit{i}) exchange position with $\ell^{\prime\prime}$, then \textit{ii}) be transported to its periodic replica, and finally \textit{iii}) exchange positions again with the replica of $\ell^{\prime\prime}$ belonging to the same cell where $\ell^{\prime}$ is found after step \textit{ii}. In this path, the nucleus $\ell^\prime$ transports a net charge equal to $q_{\ell^{\prime\prime},\alpha}$, since the first particle exchange---step \textit{i}---cancels out with the second---step \textit{iii}. At the same time, this path shares the same endpoints with the paths where only $\ell^\prime$ is moved to its periodic replica, where a charge equal to $q_{\ell^\prime,\alpha}$ is transported. Therefore, ${q_{\ell^\prime,\alpha} = q_{\ell^{\prime\prime},\alpha}}$ and the topological charge only depends upon the nuclear species, i.e.,~${q_{\ell^\prime,\alpha}=q_{S(\ell^\prime),\alpha}}$. The third condition, isotropy, is proved by first noticing that, under SA, the dipole displaced along a supercell vector is parallel to the displacement vector~\cite{jiang2012rigorous}; then, the result simply follows from additivity by equating the dipoles displaced by $L$ along any two Cartesian directions in sequence, say ${\hat{\vec{x}}=(1,0,0)}$ and ${\hat{\vec{y}}=(0,1,0)}$, to the dipole displaced along the sum of the two directions, i.e.,~${(1,1,0)}$. We conclude that ${q_{\ell,\alpha}=q_{\ell}}$ and ${q_{\ell^\prime}=q_{\ell^{\prime\prime}}}$.
In this way, we proved that the set of topological charges $\{q_\ell\}$ meet all the requirements for what chemistry would call OSs of the chemical species in the ionic conductor.

We are now in the position to prove Equation~\eqref{eq:GB theorem}, which is most compactly delivered by Resta's many-body formulation~\cite{resta2021faraday}. Let us evaluate Equation~\eqref{eq:einstein-helfand sigma} with the expression for the dipole displacement given by Equation~\eqref{eq:Delta mu as a phase difference}:
\begin{align}\label{eq:sigma einstein-helfand with delta gamma}
    \sigma = \frac{e}{6 \pi L^2 k_{\mathrm{B}} T} \lim_{t \to \infty} \frac{\expval{\abs{\bm{\Delta\gamma}(t)}^2}}{2 t}.
\end{align}
After a long enough time $t$, which is implicit in the ${t \to \infty}$ limit, the angles $\bm{\Delta\gamma}(t)$ accumulated between the initial and final configurations, respectively $\{\vec{R}_\ell(0)\}$ and $\{\vec{R}_\ell(t)\}$, will be much larger than $2\pi$. By the same token, the nuclei are expected to be in different periodic cells with respect to the one they started in. Given the final configuration, one could always imagine artificially moving each of the nuclei from $\{\vec{R}_\ell(t)\}$ to the periodic replica of $\{\vec{R}_\ell(0)\}$ in the same cell they are placed at time $t$. Let us indicate with $\bm{\delta\gamma}$ the additional phase accumulated during this last portion of trajectory; since it results from displacing all the nuclei within the same cell, it is assured that ${\delta\gamma_\alpha<2\pi}$. The trajectory defined in this way is periodic in PBCs, and therefore the transported charge is an integer multiple of $e$. From the definition of $\bm{\gamma}_{\ell}$ of Equation~\eqref{eq:gamma per nucleus}, it follows that each nucleus transports exactly $q_\ell$ elementary charges. It is precisely the same as if the phases accumulated by each nucleus had been computed as
\begin{align}\label{eq:gamma prime}
    \bm{\gamma}_\ell^\prime(t)=\frac{2\pi}{L} q_\ell \vec{R}_\ell(t),
\end{align}
their sum being denoted by $\bm{\Delta\gamma}^\prime(t)$; i.e.,~it is equivalent to computing the electrical conductivity using $\bm{\Delta\mu}^{\prime}(t)$.
The electrical conductivity computed with ${\bm{\Delta\gamma}^\prime(t) = \bm{\Delta\gamma}(t) + \bm{\delta\gamma}}$ reads
\begin{widetext}
\begin{align}\label{eq:sigma prime contributions}
    \sigma^{\prime} = \frac{e}{3 L^2 k_{\mathrm{B}} T} \lim_{t \to \infty} \frac{1}{2t} \left[ \expval{\abs{\frac{\bm{\Delta\gamma}(t)}{2\pi}}^2} + \expval{\abs{\frac{\bm{\delta\gamma}(t)}{2\pi}}^2} + 2 \expval{\frac{\bm{\Delta\gamma}(t) \cdot \bm{\delta\gamma}(t)}{4\pi^2}} \right].
\end{align}
\end{widetext}
For long times, each $\Delta\gamma_\alpha$ is much larger than $2\pi$ while each $\delta\gamma_\alpha$ is bounded by the same quantity: we conclude that the second and third terms in Equation~\eqref{eq:sigma prime contributions} do not contribute. This long-time limit may also be probed in general by inspecting whether ${\abs{\bm{\Delta \gamma}} \gg 2\pi}$. This proves the theorem of Equation~\eqref{eq:GB theorem}.

In this section, we have demonstrated that, under SA, each nucleus carries an integer charge over macroscopic distances which can uniquely be identified with the atomic OS of its species. The OS of a given chemical species is the same for every nucleus of that species; moreover, OSs are additive and independent of atomic positions. The consequence of these facts is that the electrical conductivity of the material can be computed from the dipole displacement given by the sum over nuclei of the atomic OSs times their velocities. Charge transport is purely convective; the only mechanism by which charge can be transported is mass diffusion. This transport regime is referred to as \emph{trivial}.

\subsection{Numerical experiments}

A demonstration of the quantization of the charge carried by nuclei in an emblematic liquid electrolyte is shown in Ref.~\onlinecite{grasselli2019topological} and rediscussed here. The reader is referred to that paper for technical and methodological details on the simulation. A system of molten KCl is modeled through a 64-atoms---32 per species---cubic simulation cell with a side $L=14.07\,\text{\AA}$. Nuclei are described as classical particles, while the quantum nature of electrons is included in a mean-field sense through density-functional theory (DFT) in the Perdew-Burke-Ernzerhof (PBE) flavor~\cite{perdew1996generalized}. 

\begin{figure}[tb]
    \centering
    \includegraphics[width=\columnwidth]{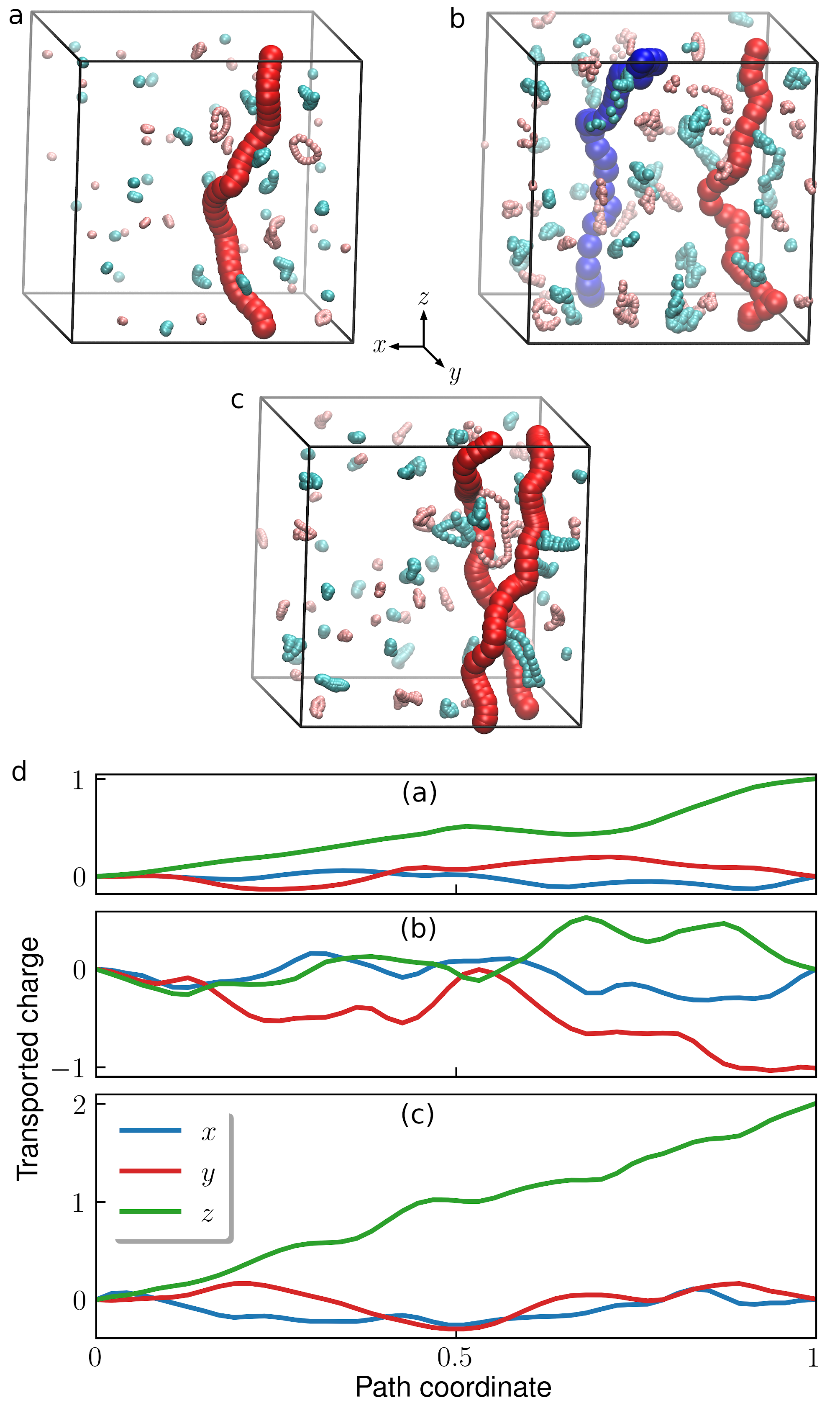}
    \caption{Pendry-Hodges \emph{Gedankenexperiment} under SA. In the simulation cell there is a stoichiometric melt of 32 K and 32 Cl atoms. K and Cl nuclei are the pink and cyan spheres, respectively. The larger spheres are the nuclei in evidence. (a-c) Periodic MEPs of a K nucleus (in red) in the $z$ direction (a); of a cation along $\hat{\vec{z}}$ (in red) and of an anion along $\hat{\vec{y}}$ and $\hat{\vec{z}}$ (in blue) (b); of two cations along $\hat{\vec{z}}$ with an interchange between the two. 
    In panel (d), the charge transported along each direction in the three paths is shown in units of the elementary charge. The path coordinate goes from 0 at the initial configuration to 1 at the final one. This Figure is adapted from Ref.~\onlinecite{grasselli2019topological}, from which the data are taken.}
    \label{fig:SA worms}
\end{figure}

A configuration is drawn from an ab initio molecular dynamics (AIMD) simulation equilibrated at $1200\,\text{K}$ and used as a reference.
From this configuration, the topological OSs are measured via an implementation of the Pendry-Hodges \emph{Gedankenexperiment} (see Section~\ref{sec:topological oss}). Different nuclei are randomly chosen and displaced from their initial position to a nearby periodic cell along minimum energy paths (MEPs), i.e.,~with the other atoms moving out of the way to minimize the total energy of the system via the nudged elastic bands method~\cite{henkelman2000a}. The displaced charge is obtained by integrating the electric flux as given by Equation~\eqref{eq:J Born}, with the Born effective charges being computed at each step of the trajectory via density-functional perturbation theory (DFPT)~\cite{baroni2001phonons}. Three versions of this experiment are carried out, as illustrated in Figure~\ref{fig:SA worms} by the MEPs in the simulation cell (top panels) and a plot of the respective transported charges (bottom panels): the displacement of a single K nucleus to one of its periodic images (Figure~\ref{fig:SA worms}a,d); the displacement of a K nucleus to its replica in direction $\hat{\vec{z}}$, and a Cl nucleus to its replica in the cell in direction ${\hat{\vec{z}}+\hat{\vec{y}}}$ at the same time (Figure~\ref{fig:SA worms}b,d); the displacements of two different K nuclei in direction $\hat{\vec{z}}$ at the same time with inversion of their final positions---i.e.,~the former ends in the periodic replica of  the initial position of the latter, and vice versa (Figure~\ref{fig:SA worms}c,d). The experiments confirm what is exposed in Section~\ref{sec:topological oss}: as one would expect from chemistry, K nuclei displace a charge equal to $+e$, while Cl nuclei $-e$, in whatever direction they are moved. This means that $q_{\mathrm{K}}=+1$, and $q_{\mathrm{Cl}}=-1$. The charge transported along the directions other than the one where motion happened are exactly zero. Furthermore, charge displacement is additive: when a pair of K and Cl atoms is moved in the same direction, there is no net charge transport. The case where the two nuclei are moved and their positions interchanged is also revealing: while each of the two paths alone is not periodic, their concatenation is, since it is the displacement of a nucleus to its periodic replica two cells apart, rather than one. Each of the two displacements would carry two elementary charges over a length of two sides of the cell, resulting in a charge of $4e$ for the cyclic path, i.e.,~$2e$ for half the path, when positions are interchanged.

\section{Polarization and Wannier Functions}\label{sec:wannier}

\begin{figure}[tb]
    \centering
    \includegraphics[width=\columnwidth]{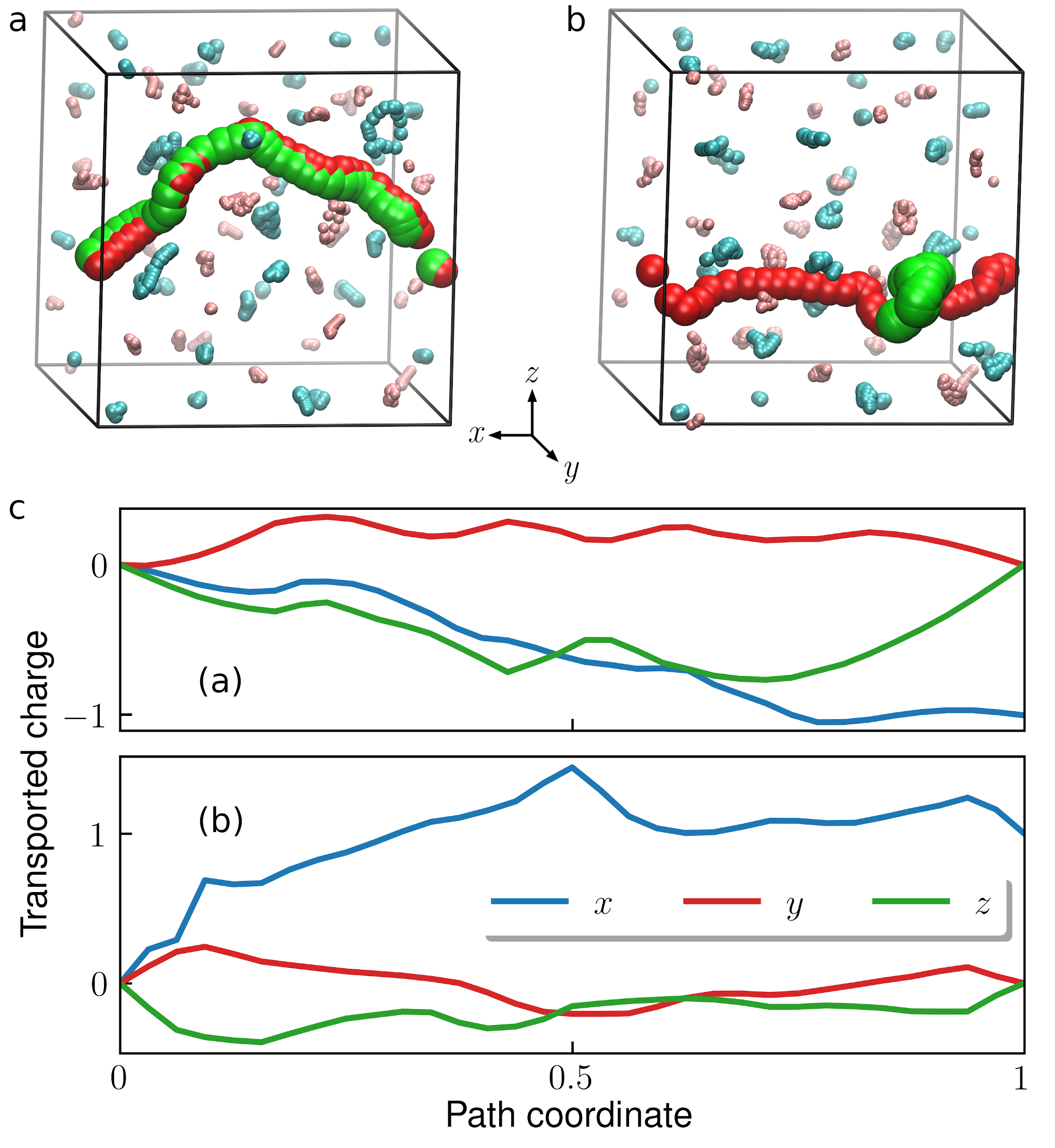}
    \caption{Pendry-Hodges \emph{Gedankenexperiment} when SA does not hold. In the simulation cell there is a nonstoichiometric melt of 33 K nuclei, depicted as pink spheres, and 31 Cl nuclei, in cyan. The green sphere represents the WC associated with the excess lone pair. The larger spheres are the particles in evidence. (a-b) Periodic MEPs of the same K nucleus (in red) in the $x$ direction, starting and ending in the same points in two nearby cells: while for one path the K nucleus drags the WC along its motion (a), in the other it does not (b). 
    In panel (c), the charge transported along each direction for the two paths is shown in units of the elementary charge. The path coordinate goes from 0 at the initial configuration to 1 at the final one. This Figure is adapted from Ref.~\onlinecite{pegolo2020oxidation}, from which the data are taken.}
    \label{fig:broken SA worms}
\end{figure}

In the case of independent (e.g., Kohn-Sham) electrons, the insulating singlet ground state of a system of $N$ nuclei and $N_{\mathrm{el}}$ electrons can be written as a Slater determinant of $N_{\mathrm{el}}/2$ doubly occupied Bloch orbitals. Equivalently, the same wavefunction can be written on a basis of WFs by means of a unitary transformation, whose positions are referred to as Wannier centers (WCs), $\{\vec{R}^{(W)}_j\}$~\cite{marzari1997maximally}. The electronic charge density is thus partitioned into localized contributions. While this partitioning is gauge-dependent, i.e.,~both WFs and WCs are nonunique, the sum of the WCs in the central cell is gauge-independent, and it happens to be  equal to the total polarization~\cite{vanderbilt2018berry}. The total electric current, i.e.,~\cite{pegolo2020oxidation, resta2021faraday}
\begin{align}\label{eq:J wannier}
    \vec{J}(t) = \frac{e}{L^3} \left[\sum_{\ell=1}^{N} Z_{\ell} \dot{\vec{R}}_{\ell} - 2 \sum_{j=1}^{N_{\mathrm{el}}/2} \dot{\vec{R}}^{\mathrm{(W)}}_j \right],
\end{align}
assumes a classical-like expression. In this picture, the Pendry-Hodges \emph{Gedankenexperiment} assumes the intuitively clear meaning: a  nucleus, when displaced along a direction of the torus, drags some of the Kohn-Sham WFs, thus transporting an even number of electrons. The resulting OS is the bare charge of the nucleus minus twice the number of transported WFs~\cite{pegolo2020oxidation, resta2021faraday}. The situation can also be clearly visualized by looking at the trajectory of the nuclei and the WCs in the central cell (see also Figure~\ref{fig:broken SA worms} and Refs.~\onlinecite{jiang2012rigorous} and~\onlinecite{pegolo2020oxidation}).

\section{Breakdown of strong adiabaticity}\label{sec:breakdown of SA}

In nature, chemically relevant situations occur where different atoms of the same species feature different OSs depending, e.g., on the local chemical environment.

There are circumstances where a meaningful topological definition of OSs is still possible. This is the case exemplified by Figure~\ref{fig:ACS zoology}b: strongly adiabatic domains are completely separated by metallic regions. OSs can be uniquely assigned to nuclei within the same domain, and the properties of OSs discussed above remain valid. For instance, in Ref.~\onlinecite{jiang2012rigorous} the authors found that two OSs for the same atomic species (bismuth) coexist in BaBiO\textsubscript{3}, where Bi atoms in the octahedral sites feature OSs equal to ${+3}$ or ${+5}$, depending on their distance to the O atoms placed in the vertices of the octahedra.
Nonetheless, we remark that the exchange of the two Bi atoms \emph{cannot} be possible without crossing a metallic region. The same mechanism is present in ferrous-ferric aqueous solutions~\cite{hudis1956rate}, where the charge transfer turning Fe(II) into Fe(III) and vice versa can only be accomplished via a nonadiabatic charge transfer.
Another remarkable case where the SA of an insulating system is globally  dropped, but the OSs of its nuclei can still be defined, is represented by the insulating subsystem of metal-insulator interfaces~\cite{stengel2007ab}, where the system is metallic along some Cartesian directions and nonmetallic along others.

A completely different situation occurs when the breakdown of SA is accompanied by the presence of metallic regions that can be encircled by closed adiabatic paths, as pictured in Figure~\ref{fig:ACS zoology}c~\cite{pegolo2020oxidation}. 
In this case, charge transport can no longer be topologically classified in terms of the number of cells spanned by each nucleus along each direction, encoded in $\vec{n}$, and the very concept of OS loses much of its topological meaning. In fact, there can be trivial loops in the ACS (i.e.,~with $\vec{n}=0$) that cannot be shrunk to a point without ever crossing a metallic region. This amounts to saying that $\bm{\mathfrak{z}}$ (or some Cartesian component thereof) rotates once around the origin of the complex plane, where the system is metallic, even if the configuration does not move to a different periodic replica in the ACS. Thouless' theorem can still be leveraged to conclude that the dipole displaced along such a path is quantized but, at variance with the strongly adiabatic case, it is possibly nonzero, whereas the atomic net displacements vanish because ${\vec{n}=0}$. 

The repercussion of this paradigmatic situation to macroscopic electrical conduction is that charge transport is no more completely correlated to mass transport. This transport regime is somehow intermediate between insulating and metallic behaviors: while the system is always insulating and the charge motion is uniquely dictated by the adiabatic motion of the ions, in this situation a net charge transfer is possible even without the need for a net mass diffusion; we call this regime \emph{nontrivial transport}~\cite{pegolo2020oxidation}. 

\subsection{Numerical experiments}

One of the hallmarks of the breakdown of SA is the possibility to pump charge without any net mass displacement. In Ref.~\onlinecite{pegolo2020oxidation} this is demonstrated via two toy-model systems---namely, the linear H\textsubscript{3}\textsuperscript{+} cation in PBCs and the neutral K\textsubscript{3}Cl complex. In these simple cases, it is easy to locate (and visualize) the regions of the ACS where the system is gapless.
Nuclear trajectories can then be devised to be loops that encircle such a region, without ever crossing it. In both these toy-model systems, the result---the pumping of $-2$ electronic charges---confirms that an electron pair is transported even if the nuclei return to their initial position in the original cell, and no net ionic displacement occurs. 

Besides these two proofs of concept, another clear example of the breaking of SA is given in Ref.~\onlinecite{pegolo2020oxidation} for the more realistic case of the nonstoichiometric molten KCl with additional K atoms. This system is modeled through the same simulation cell as for the stoichiometric case, but with an unbalanced proportion of K and Cl atoms; i.e.,~33 K atoms and 31 Cl atoms. An AIMD simulation of this nonstoichiometric molten salt is carried out, and the system is checked to stay insulating during the whole dynamics---in an independent electron framework, this means that the spectral gap stays open.

We repeat the Pendry-Hodges experiment on a configuration drawn from the K\textsubscript{33}Cl\textsubscript{31} AIMD simulation~\cite{pegolo2020oxidation}.
A typical configuration is chosen as a reference; a single K nucleus is driven along two different paths to its periodic replica in an adjacent cell, the other nuclei being let free to adjust their position at each step of the trajectory. The measured topological charges are not the same along the two paths: they are equal to $+e$ in one case, and to $-e$ in the other~\cite{pegolo2020oxidation}. This means that the two paths cannot be continuously deformed into one another without crossing a region where the gap closes. In other words, SA is broken, and topological OSs are ill-defined. 

Of course, one cannot expect the dynamics of a many-body interacting system to return to its initial position in any finite time. What is observed in physical AIMD simulations, however, is the presence of an electron pair, dissolved among the ions, that is rather \emph{free to diffuse}. This phenomenon---the formation of localized electron pairs---is often referred to as \emph{bipolarons}~\cite{selloni1987electron, fois1988bipolarons, selloni1989simulation, fois1989approach}. A single bipolaron is found to contribute the most to the electrical conductivity of K\textsubscript{33}Cl\textsubscript{31}, which turns out to be almost 5 times larger than the one of the stoichiometric melt~\cite{pegolo2020oxidation}. Notably, the bipolaronic current is largely uncorrelated from the ionic one. This confirms that the breakdown of SA allows adiabatic charge transport without the need for mass convection.

\section{Charge-transfer reactions}\label{sec:charge transfer}

\begin{figure}
    \centering
    \includegraphics[width=\columnwidth]{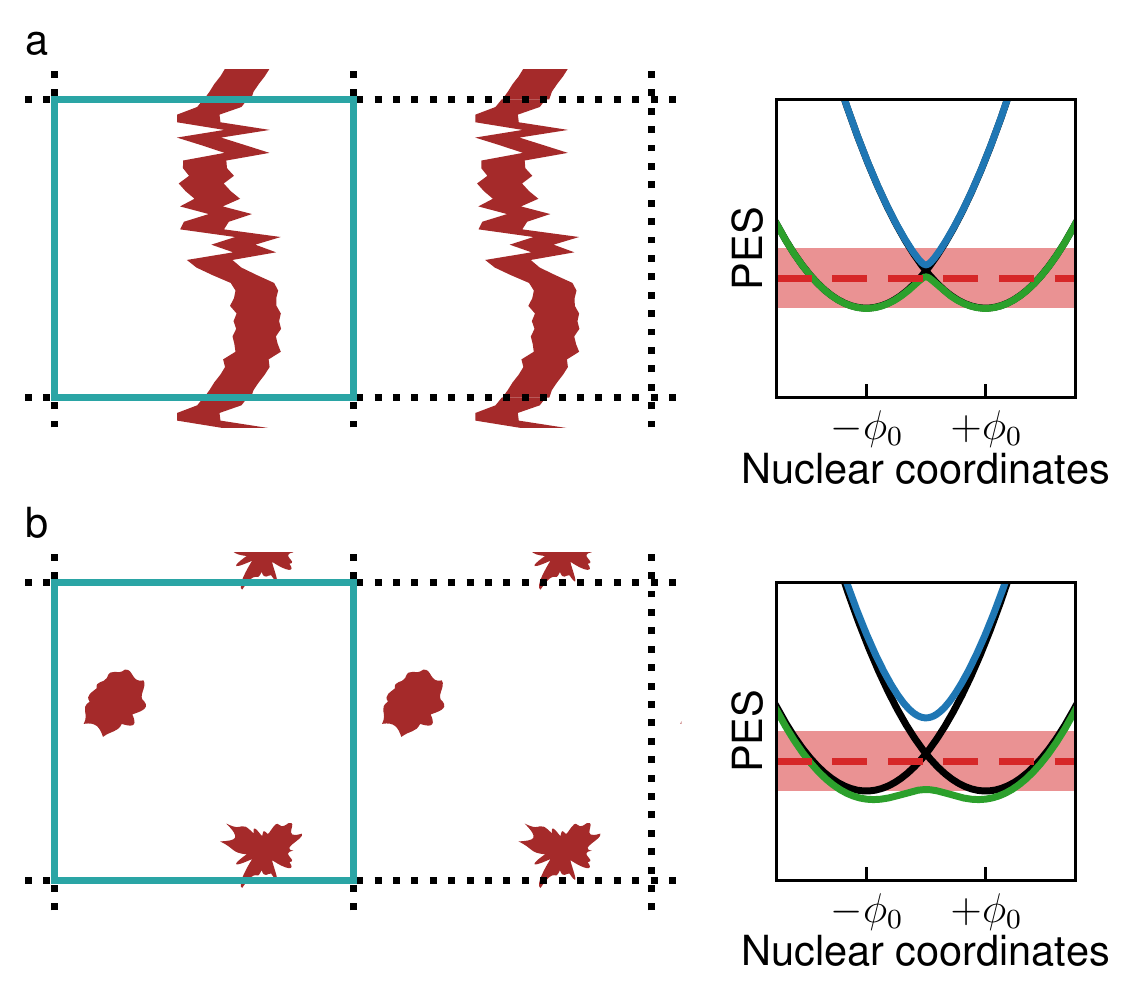}
    \caption{Parallel between our topological theory and Marcus-Hush ET theory. (a) ACS with strongly adiabatic domains separated by metallic regions (left) and PESs with $p \gg 1$, signaling nonadiabatic charge transfer (right). (b) ACS where SA is broken (left) and PESs with ${p \gtrsim 1}$, where adiabatic charge transfer is possible.}
    \label{fig:all-Marcus}
\end{figure}

Electron transfer (ET) reactions are a class of chemical reactions that involve the exchange of charge between compounds, i.e.,~from a reactant state to a product state. The first successful attempts to provide a theory of electrochemical reactions are due to Marcus~\cite{marcus1956on} and Hush~\cite{hush1958adiabatic}. In their seminal works, they were able to address the matter of outer-sphere electron transfer---i.e.,~where the reacting species remain separated during the whole process---with a particular focus on the role of the solvent reorganization, rather than a detailed account of electronic interactions which, at the time, could not be included in a quantitative treatment due to the complex nature of the calculations involved.
We summarize below the main elements of this theory.

A typical Marcus-Hush scenario involves two reacting partners in proximity. There usually is a free energy barrier separating reactants and products. In fact, the initial and final states bear different charges, and reactant and product states are solvated. Therefore, a reorganization of the environment is required for the electron to overcome the barrier and the ET to happen. The situation can be understood as a two-state system~\cite{weiss2012quantum}: a reactant and a product state. 
A simple way to incorporate the effect of the solvent is through a linearly responding heat bath. The Hamiltonian which represents this system is the spin-boson Hamiltonian, i.e.,
\begin{align}\label{eq:spin boson}
    \hat{H} = -\frac{1}{2}\hbar \Delta \hat{\varsigma}_x -\frac{1}{2}\hbar \epsilon \hat{\varsigma}_z + \frac{1}{2}\mu\mathcal{E}(t)\hat{\varsigma}_z+\hat{H}_R,
\end{align}
where $\{\hat{\varsigma}_{\alpha}\}$ are the Pauli matrices; $\Delta$ and $\epsilon$ are the hopping rate and on-site bias, respectively; $\mu\mathcal{E}(t)$ describes the collective bath mode coupled to the electronic system and can be thought of as a fluctuating dynamical polarization energy due to the local environment~\cite{weiss2012quantum}, and its properties are entirely contained in its correlation function; $\hat{H}_R$ is the Gaussian reservoir associated to the solvent. When the bath is purely adiabatic, which is the condition we are interested in, the bath correlation function reduces to a constant, $2 k_\mathrm{B} T \Lambda$, where $\Lambda$ can be understood as the reorganization energy of the solvent. The contribution of the bath to the Hamiltonian in Equation~\eqref{eq:spin boson} plays no role, and what remains reduces to (in matrix form)~\cite{carmeli1985effective}:
\begin{align}\label{eq:spin boson adiabatic}
    \hat{H} = \frac{\hbar}{2}\left(
\begin{array}{cc}
 \frac{ \mu\mathcal{E}}{\hbar }-\epsilon  & -\Delta  \\
 -\Delta  & \epsilon -\frac{\mu\mathcal{E}}{\hbar } \\
\end{array}
\right).
\end{align}
The eigenstates of $\hat{H}$---i.e.,~the electronic adiabatic states---are easily found to have eigenvalues
\begin{align}
    E_{\pm}(\mathcal{E}) = \pm \frac{\hbar}{2} \sqrt{\Delta^2 + (\epsilon-\mu\mathcal{E}/\hbar)^2}.
\end{align}
{Averaging over} the bath fluctuations, one obtains the adiabatic potential energy surfaces (PES), $F_{\pm}$, as a function of the mean polarization energy $\mu\mathcal{E}$~\cite{weiss2012quantum}:
\begin{align}\label{eq:adiabatic potential energy}
    F_{\pm}(\mathcal{E}) = \frac{\mu^2 \mathcal{E}^2}{4\Lambda} \pm \frac{\hbar}{2} \sqrt{\Delta^2 + (\epsilon - \mu\mathcal{E}/\hbar)^2}.
\end{align}
When there is no on-site bias, i.e.,~reactant and product are solvated with the same energy, $\epsilon=0$. The adiabatic surfaces are qualitatively different according to the parameter $p=\Lambda/(\hbar\Delta)$, where the electronic coupling $\hbar\Delta$ can be thought of as the unit of energy. Doing so also for the polarization energy, i.e.,~defining $\phi=\mu\mathcal{E}/(\hbar\Delta)$, Equation~\eqref{eq:adiabatic potential energy} becomes
\begin{align}\label{eq:adiabatic potential energy reduced}
    F_{\pm}(\phi) = \frac{\hbar \Delta}{2} \left( \frac{\phi^2}{2p} \pm \sqrt{1+\phi^2} \right).
\end{align}
The reactant and product states are located at $\mp\phi_0$, respectively, where ${\phi_0=\sqrt{p^2-1}}$.
In the adiabatic limit, the electronic coupling is so large that $p$ is of order 1 or lower. When $p<1$ there are no environmental fluctuations; both surfaces feature a single minimum, therefore no electron is transferred. For $p \gtrsim 1$, there are two minima in the ground adiabatic surface, separated by a barrier whose height depends on $p$. The electronic coupling is still large, but there are sufficient environmental fluctuations to allow an electronic transfer. At the opposite limit, $p \gg 1$, there is the nonadiabatic transfer, when the fluctuations are so large with respect to the electronic coupling that the gap between $F_-$ and $F_+$ is very small, and the electron has a large probability of jumping to the higher energy surface, invalidating the Born-Oppenheimer approximation.

The two regimes $p \gtrsim 1$ and $p \gg 1$ outlined above where ET can happen can be understood from our perspective as pictured in Figure~\ref{fig:all-Marcus}. In the two panels, on the left there is a sketch of the ACS with the central cell circled in light blue, and a periodic replica on its right, like those in Figure~\ref{fig:ACS zoology}. On the right there are the PESs---the black ones are the Marcus parabolas representing diabatic states, the green one is the adiabatic ground state, and the blue one is the adiabatic excited state. The red horizontal line represents the average energy of the system, the shaded area being the entity of thermal fluctuations, i.e.,~of the order $2k_\mathrm{B}T$. 

In panel (a), the ACS features metallic walls that cannot be bypassed. OSs are well defined only within a given adiabatic domain and, for a given nuclear species, they can be different for different adiabatic domains. Charge can be transferred only by crossing a metallic region, i.e.,~by breaking adiabaticity. Any electron transfer is nonadiabatic: since the electronic coupling between reactant and product states is too low with respect to the reorganization energy, $p \gg 1$, and the adiabatic energy surfaces practically coincide with the diabatic ones. 
This phenomenon is observed in physical system where atoms of the same species feature different OSs, such as the paradigmatic case of the ferrous-ferric exchange in water~\cite{hudis1956rate},
\begin{align}
    \mathrm{Fe^{2+}(aq)}+\mathrm{Fe^{3+}(aq)} \rightleftharpoons \mathrm{Fe^{3+}(aq)}+\mathrm{Fe^{2+}(aq)}.
\end{align}
For the exchange to happen, the system must pass through a point where the electronic levels are degenerate, and adiabaticity is lost. 
Panel (b) displays an ACS where SA is broken and there are metallic regions that can be encircled by adiabatic loops.
Therefore, there exist adiabatic paths, sharing the same endpoints, that cannot be deformed into one another without ever crossing a metallic region, as well as trivial loops in the ACS which can pump an integer charge to another cell without any net displacement of the nuclei. As we proved in Section~\ref{sec:breakdown of SA}, this implies that a topological, unique definition for the atomic OSs is not admitted, and the \emph{adiabatic} motion of an erratic, localized electronic charge can be observed.
In the Marcus picture, this is exactly what happens in the intermediate adiabatic regime, $p \gtrsim 1$. Electron transfer can happen as an effect of nuclear fluctuations within the Born-Oppenheimer approximation, since the adiabatic surfaces are sufficiently separated. This is observed, e.g.,~in nonstochiometric molten salts, where charge is mainly transported by localized electronic charge (polarons or lone pairs) diffusing through the system~\cite{pegolo2020oxidation} by means of an activated Marcus process mediated by the nuclear thermal motion.

\section{Applications}\label{sec:applications}

Besides giving a solid framework for the topological classification of ab initio charge transport in electronic insulators, the theory exposed so far has direct applications to the simulation of charge transport via AIMD. In fact, by formally justifying the use of OSs in the calculation of the electric charge fluxes entering the Green-Kubo formula, our theory dramatically reduces the computational load needed to extract the ab initio $\sigma$, since it avoids the cumbersome task of computing Born effective charge tensors for each atom, along the simulation. 
In the first part of this Section molten salts are used to validate the general method and elucidate the differences between SA and its breaking; we then show some relevant applications in the fields of planetary and energy materials, where the use of integer charges in the definition of the electric flux has been successfully employed.

\subsection{Validation of the method}

\begin{figure}
    \centering
    \includegraphics[width=\columnwidth]{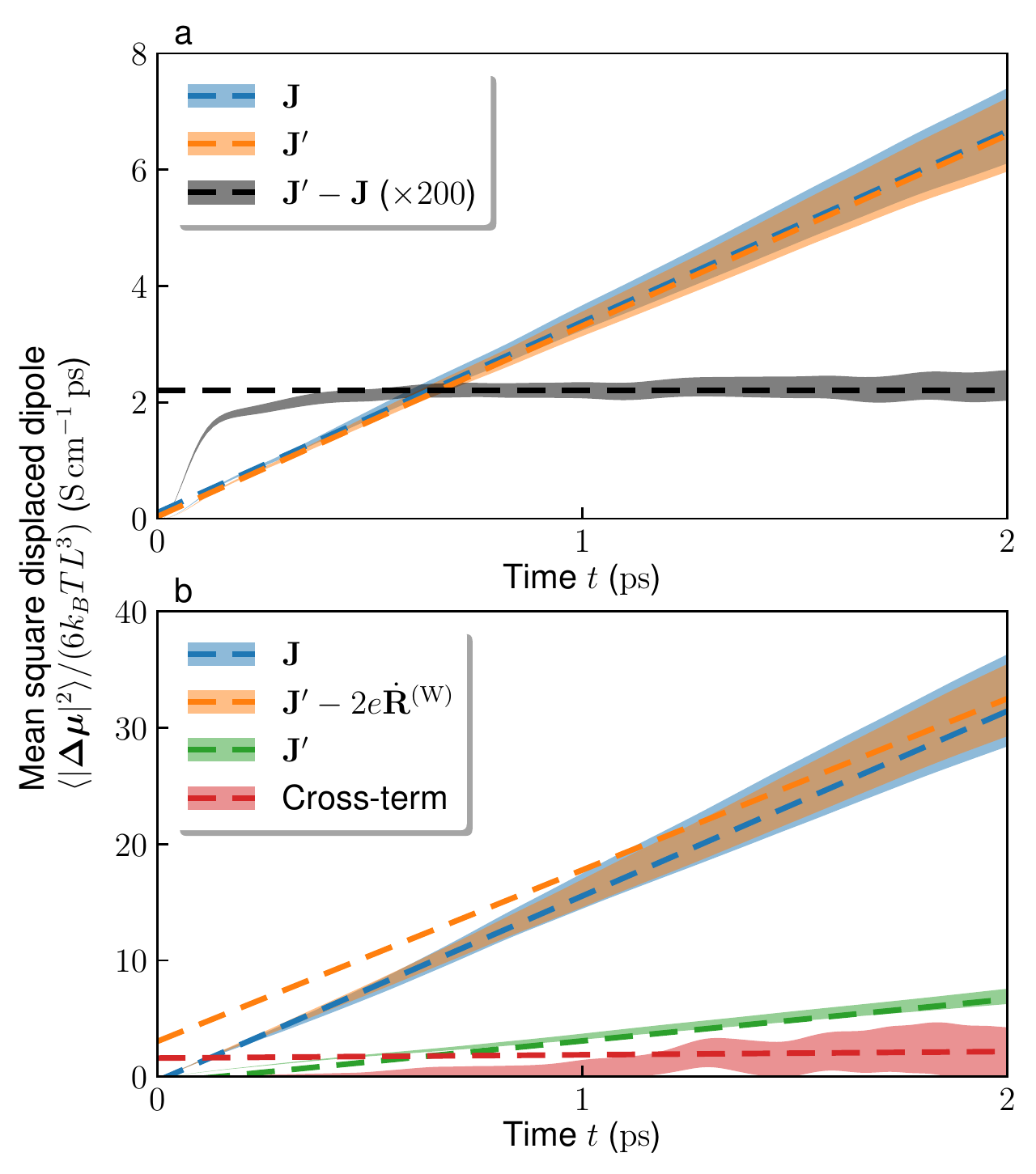}
    \caption{MSDDs of molten salts computed with different definitions of the electric flux. (a) The case of stoichiometric KCl: both the exact electric flux and the topological flux yield the same electrical conductivity. This picture is adapted form Ref.~\onlinecite{grasselli2019topological}. (b) The case of nonstoichiometric K-KCl melt: the topological flux plus the lone pair contribution are equivalent to the exact definition of the electric flux. The topological flux alone yields a value of $\sigma$ much lower than the correct one. The cross-correlation term is compatible with zero: the ionic and lone-pair contributions are uncorrelated to one another. This picture is adapted from Ref.~\onlinecite{pegolo2020oxidation}.}
    \label{fig:molten salts msqd}
\end{figure}

A numerical experiment on the validity of Equation~\eqref{eq:GB theorem} was given in Ref.~\onlinecite{grasselli2019topological} for molten KCl. The electrical conductivity was computed from the time-series of the electric flux, both the ab-initio one, $\vec{J}$, given by Equation~\eqref{eq:J Born} and employing Born effective charge tensors, and the ``topological'' flux, $\vec{J}^\prime$, where charges are the topological OSs computed as described in Section~\ref{sec:topological oss}. The values of $\sigma$ and $\sigma^\prime$ were found to be $3.2\pm0.2\,\mathrm{S\,cm^{-1}}$ in both cases, so that the two formul\ae~perfectly agree, even on the statistical uncertainty of the measure. A plot of the mean square displaced dipole (MSDD), whose slope yields the value of the electrical conductivity (see Equation~\eqref{eq:einstein-helfand sigma}, is shown in Figure~\ref{fig:molten salts msqd}a. The reported values have been also checked with a cepstral analysis~\cite{ercole2017accurate, marcolongo2020gauge}, implemented in \textsc{SporTran}~\cite{ercole2022sportran, ercole20172022textsc}.
The  MSDD of the difference of the fluxes given by the two formul\ae~is also shown: its slope is compatible with zero.

The nonstoichiometric molten salt K\textsubscript{$x$}(KCl)\textsubscript{$1$$-$$x$}, with $x \simeq 0.6$, was treated in Ref.~\onlinecite{pegolo2020oxidation}. There, the electrical conductivity was estimated both by computing the electric flux as the time-derivative of the total polarization (equivalent to Equation~\eqref{eq:J Born}), and by using the flux given by the sum of the topological flux and the contribution due to the diffusing electronic lone pair, $\vec{J}^{\prime\prime}$, i.e.:
\begin{align}
    \vec{J}^{\prime\prime} = \vec{J}^\prime - 2e \dot{\vec{R}}^{(W)}_{\mathrm{lp}},
\end{align}
where $\vec{R}^{(W)}_{\mathrm{lp}}$ is the WC of the (doubly occupied) WF associated with electronic lone pair (hence the subscript $\mathrm{lp}$). The electronic lone-pair contribution is the most relevant to the value of electrical conductivity, as shown in Figure~\ref{fig:molten salts msqd}b. Moreover, ionic and lone-pair contributions are found to be largely uncorrelated to one another, as evidenced by the cross-correlation term (in red in Figure~\ref{fig:molten salts msqd}b), $-2e\langle \bm{\Delta}\vec{R}^{(W)}_{\mathrm{lp}} \cdot \bm{\Delta}\bm{\mu}^\prime \rangle$, which has a vanishing asymptotic slope.
This is a macroscopic manifestation of the fact that charge can be transported even without a net mass displacement.

\subsection{Water at extreme conditions}

Water at high-$pT$ conditions is a major constituent of celestial bodies formed far enough from their host star for water to condense~\cite{lodders2003solar}. In the mantle of ice giants, like Uranus and Neptune, water is present as a partially dissociated conducting liquid, which is thought to be responsible for the magnetic fields of these celestial bodies via a dynamo mechanism~\cite{kivelson1996discovery}. 
At typical pressures and temperatures of interest of the outer core, instead, (e.g., $240\,\mathrm{GPa}$ and $5000\,\mathrm{K}$ at half the radius of Uranus), water is in the super-ionic phase, i.e.,~the oxygen ions occupy the sites of a crystalline lattice, while protons are free to diffuse in liquid-like fashion~\cite{cavazzoni1999superionic, millot2018experimental}.

Detailed knowledge of the transport properties of different phases of H\textsubscript{2}O occurring at high-$pT$ conditions is key to any quantitative evolutionary model of water-rich celestial bodies. In spite of the steady progress in diamond-anvil-cell and shock-wave technologies, the experimental investigation of transport properties of materials at planetary conditions is still challenging~\cite{millot2018experimental}. In the specific case of H\textsubscript{2}O, the electrical conductivity is only known with large uncertainties along the Hugoniot curve on a limited portion of the $pT$ diagram~\cite{mitchell1982equation, yakushev2000electrical}. Computer simulations may be our only handle on the properties of matter at physical conditions that cannot be achieved in the laboratory. In particular, AIMD simulations allow to accurately simulate the bulk properties of super-ionic water~\cite{french2011dynamical, french2016ab}. 
The electrical conductivity can be obtained at the same computational cost of an equilibrium AIMD simulation thanks to Equation~\eqref{eq:GB theorem}, i.e.,~by using the electric flux defined as
\begin{align}\label{eq:J H2O}
    \vec{J}^\prime(t) = -\frac{2e}{L^3} \sum_{\ell \in \mathrm{O}} \dot{\vec{R}}_\ell + \frac{e}{L^3} \sum_{\ell \in \mathrm{H}} \dot{\vec{R}}_\ell
\end{align}
instead of the true electric flux (Equation~\eqref{eq:J Born}) which would require the computationally demanding on-the-fly calculation of Born effective charge tensors~\cite{grasselli2020heat}.

In Refs.~\onlinecite{grasselli2020heat} and~\onlinecite{stixrude2021thermal} Eq.~\eqref{eq:J H2O} was exploited to characterize the electrical conductivity of different phases of water which are expected to be present in the icy giants Uranus and Neptune. Also in this case, the fully quantum-mechanical (DFPT)~\cite{giannozzi2009quantum,giannozzi2017advanced,giannozzi2020quantum} calculation of the charge flux gives an electrical conductivity which coincides with that obtained via $\vec{J}^\prime(t)$ in Equation~\eqref{eq:J H2O}.

This procedure has been also successfully applied to the superionic phases of nanoconfined water~\cite{kapil2021the}.

\subsection{Solid-state electrolytes}

Solid-state electrolytes (SSEs) are attractive materials for Li-ion batteries as they in principle could allow to design and build safer batteries. Ionic conductivity is one of the most desired properties for this class of materials. 
When ionic correlations are negligible, as in the case of ``dilute'' systems, the Nernst-Einstein approximation is good enough to allow to estimate $\sigma$ without resorting to the computation of Born effective charges. In general, however, there might be sufficiently strong correlations among the mobile ions and the fixed matrix or even more than one diffusing species. For all these reasons, the calculation of $\sigma$ through the GK formula, which is in principle correct, is the only one that yields reliable results. The use of integer topological OSs allows also in this case to significantly reduce the computational demand of the estimation of transport properties in SSEs~\cite{gilardi2020li, materzanini2021high, pegolo2022temperature}.

\section{Conclusions}\label{sec:conclusions}

\begin{figure}
    \centering
    \includegraphics[width=\columnwidth]{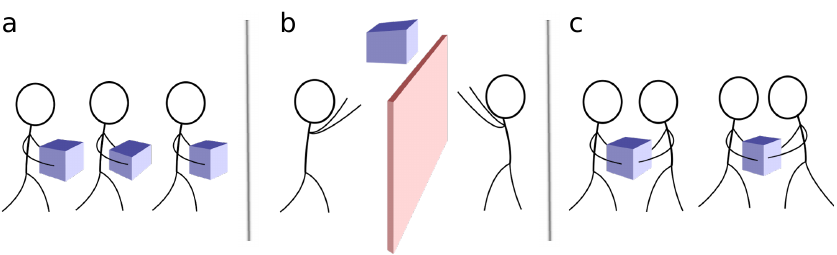}
    \caption{A pictorial description of the three charge transport mechanisms described in this review. The ``dockers'' represent the atoms, while the boxes the charge they carry. (a) Trivial transport: SA holds everywhere, each ion carries its own charge, i.e., its OS, and charge transport cannot occur without a net mass displacement. (b) The presence of metallic ``walls'' allows charge transport without mass displacement, but the price to pay is that these charge transfers are \emph{nonadiabatic}. (c) Nontrivial adiabatic transport: SA does not hold, and charge can be \emph{adiabatically} transported without a net ionic displacement, just like for dockers exchanging their boxes without going around.}
    \label{fig:scaricatori}
\end{figure}

In this article we presented a review of recent advances in the theoretical description of charge transport in electronically insulating materials. We have shown that topological quantization of adiabatic charge transport and the gauge-invariance principle of transport coefficients can be combined to give a profound insight in the mechanisms of charge transport in these materials. In fact, we are able to group these mechanisms in three main classes, depending on the topological properties of the adiabatic space, i.e., the atomic configuration space (ACS) deprived of the regions at which the system becomes metallic. These three classes are pictorially described in Figure~\ref{fig:scaricatori}, where the atoms are represented by ``dockers'', while the charge is represented by the boxes they carry. Figure \ref{fig:scaricatori}a depicts the case where strong adiabaticity (SA) holds on the entire adiabatic space: each atom/ion (effectively) carries its own charge, its OS, which is the same for atoms of the same species. In this situation, charge transport is additive and can only be accompanied by a net (macroscopic) displacement of the massive atoms. Figure \ref{fig:scaricatori}b, instead, represents the case where the global adiabiatic space is subdivided into SA domains, separated by metallic ``walls'': for each domain, SA holds, each atom can still be labeled with an OS, additivity holds and charge can only be transported with a net displacement of the nuclei; nevertheless, atoms of the same species belonging to different domains are no longer required to possess the same OS. In fact, atoms of different domains can ``pass'' charge among themselves, but only through a nonadiabatic electron transfer; only in this case can charge transfer occur without a net ionic displacement. Finally, Figure~\ref{fig:scaricatori}c depicts the case where SA is broken: it is no longer possible to uniquely assign OSs to the atoms through a formal topological procedure, like in the Pendry-Hodges \emph{Gedankenexperiment}. In this situation, which is typical, e.g., of nonstoichiometric, yet nonmetallic, molten salts, an \emph{adiabatic}, macroscopic charge transport can occur without a net atomic displacement. This is pictorially represented by the dockers exchanging their burden among themselves, but staying in their spot, without a net displacement.

\medskip

\medskip
\noindent \textbf{Data availability}\par
\noindent No new data were created or analyzed for this study.

\medskip
\noindent \textbf{Conflict of interests}\par
\noindent There are no conflicts of interests to declare.

\medskip
\noindent\textbf{Acknowledgements} \par 
\noindent This work was partially funded by the EU through the \textsc{MaX} Centre of Excellence for supercomputing applications (Project No. 824143) and by the Italian Ministry of Research and education through the PRIN 2017 \emph{FERMAT} grant. FG acknowledges funding from the European Union's Horizon 2020 research and innovation programme under the Marie Sk\l{}odowska-Curie Action IF-EF-ST, grant agreement No.~101018557 (TRANQUIL).
The authors thank Federico Berti for a critical reading of the manuscript.

\medskip

\bibliography{main}

\end{document}